\documentstyle[preprint,aps,psfig,bezier]{revtex}
\begin{document}
\tightenlines
\draft
\preprint{
\parbox{4cm}{
\baselineskip=12pt
KEK-TH-750\\ 
March, 2001\\
\hspace*{1cm}
}}
\title{Vacuum structure of spontaneously broken \\ 
${\cal N}=2$ supersymmetric gauge theory}  
\author{ Masato Arai $^a$ 
  \thanks{e-mail: m-arai@phys.metro-u.ac.jp}
     and Nobuchika Okada $^b$ 
  \thanks{e-mail: okadan@camry.kek.jp}}  

\address{$^a$Department of Physics, Tokyo Metropolitan University,\\
         Hachioji, Tokyo 192-0397, Japan}
\address{$^b$Theory Group, KEK, Tsukuba, Ibaraki 305-0801, Japan}
%
\maketitle
%
%
\vskip 2.5cm
\begin{center}
{\large Abstract}
\vskip 0.5cm
\begin{minipage}[t]{14cm}
\baselineskip=19pt
\hskip4mm
We analyze the vacuum structure of spontaneously broken 
${\cal N}=2$ supersymmetric gauge theory 
 with the Fayet-Iliopoulos term. 
Our theory is based on the gauge group $SU(2) \times U(1)$ 
 with $N_f=1,2$ massless quark hypermultiplets having 
 the same $U(1)$ charges. 
In the classical potential, there are degenerate vacua 
 even in the absence of supersymmetry. 
It is shown that this vacuum degeneracy is smoothed  out, 
 once quantum corrections are taken into account. 
In $N_f=1$ case, the effective potential is found to be 
 so-called runaway type, and there is neither well-defined vacuum
 nor local minimum.
On the other hand, in $N_f=2$ case, 
 while there is also the runaway direction in the effective potential, 
 we find the possibility that there appears the local minimum 
 with broken supersymmetry at the degenerate dyon point. 
%
\end{minipage}
\end{center}
\newpage
\section{Introduction}

There has been much progress in understanding the dynamics
 of strongly coupled ${\cal N}=1,2$ supersymmetric (SUSY) gauge theories. 
The exact effective superpotential can be derived 
 for ${\cal N}=1$ SUSY QCD (SQCD) 
 by using holomorphy properties of 
 the superpotential and the gauge kinetic function \cite{Seiberg}. 
Seiberg and Witten derived the exact low energy Wilsonian effective action 
 for ${\cal N}=2$ SUSY $SU(2)$ Yang-Mills theory \cite{s-w1}, 
 and generalized their discussion to the case 
 with up to four massive quark hypermultiplets \cite{s-w2}. 
The key ingredients which allow us to derive the exact results
 are duality and holomorphy. 
One can write down the prepotential and the gauge couplings 
 in terms of the meromorphic differential on the Riemann surface with
 genus one whose properties are determined by the dynamical scale 
 and the hypermultiplet masses.  

The results by Seiberg and Witten were extended to the case 
 with the explicit soft SUSY breaking terms by using the spurion technique. 
Unless these terms do not change the holomorphy and duality properties 
 of the theory, we can derive the exact effective action 
 for ${\cal N}=1$ and ${\cal N}=0$ (non-supersymmetric) SUSY gauge theories 
 up to the leading order for the soft SUSY breaking terms. 
In Refs.~\cite{peskin,hsu}, the exact superpotential and 
 the phase structure in ${\cal N}=1$ SQCD were discussed 
 based on the ${\cal N}=2$ SUSY gauge theory 
 with some soft breaking terms. 
In Refs.~\cite{marino1,marino2,marino3}, the vacuum structure 
 of non-SUSY gauge theory was investigated 
 in which soft SUSY breaking terms directly break 
 ${\cal N}=2$ SUSY to ${\cal N}$=0. 
As further extensions, the method to introduce non-holomorphic
 soft SUSY breaking terms was recently discussed \cite{rata}. 

In this paper, we study a spontaneously broken ${\cal N}=2$ 
 SUSY gauge theory. 
It is well known that, in the framework of ${\cal N}=2$ SUSY theory, 
 the only possibility to break SUSY spontaneously 
 is to introduce the Fayet-Iliopoulos (FI) term \cite{fayet}.
Therefore, in the following, we consider the gauge theory which includes
 $U(1)$ gauge interaction together with the FI term. 

The simplest example of this type of theory is 
 ${\cal N}=2$ SUSY QED (SQED) with the FI term \cite{arai}. 
At the classical level, although SUSY is spontaneously broken
 in Coulomb branch, 
 there are degenerate vacua (moduli space) 
 which are parameterized by the vacuum expectation value 
 of the scalar field, $a$, in the $U(1)$ vectormultiplet. 
The direction of this vacuum degeneracy in the absence of SUSY 
 is called ``pseudo flat'' direction. 
However, it is expected that this direction is lifted up, 
 once quantum corrections are taken into account. 
By virtue of ${\cal N}=2$ SUSY, the effective action is found to be 
 one loop exact, and the effective gauge coupling is given by 
 $e(a)^2\sim 1/\log (\Lambda_L/ a )$,
 where $\Lambda_L$ is the Landau pole. 
Note that there are two singular regions in moduli space, namely, 
 the ultraviolet region such as $|a| \geq \Lambda_L$, 
 and the massless singular point at the origin $a=0$. 
Since the effective potential is described as $V\sim e(a)^2$, 
 the potential minimum emerges at the origin, 
 where SUSY is formally restored. 
However, since this point is the singular point, 
 we conclude that there is no well-defined vacuum in this theory. 

In this paper, we investigate the vacuum structure 
 of more interesting theory with spontaneous ${\cal N}=2$ SUSY breaking. 
Our theory is based on the gauge group $SU(2)\times U(1)$ 
 with $N_f=1,2$ massless quark hypermultiplets 
 having the same $U(1)$ charges.  
In the ultraviolet region, the behavior of the effective potential 
 can be well understood based on the perturbative discussion, 
 since the $SU(2)$ gauge interaction is weak there. 
On the other hand, it is expected that the behavior of the effective 
 potential in the infrared region is drastically changed compared 
 with SQED, because of the presence of the $SU(2)$ gauge dynamics.

The paper is organized as follows. 
In the next section, we briefly discuss the classical structure 
 of our theory. 
It is shown that the classical potential has the pseudo flat direction. 
In Sec.~III, the low energy effective action is discussed. 
In the subsection A, we first make our framework clear, 
 and give general formulae of the effective action. 
The effective potential can be read off from this effective action, 
 and is explicitly presented in the subsection B. 
In the subsection C, we give the explicit formulae 
 for the periods and the effective gauge couplings 
 which are necessary to analyze the effective potential. 
In Sec.~IV, the effective potential is numerically analyzed, 
 and the vacuum structures of our theory are investigated 
 for both cases of $N_f=1$ (subsection A) and $N_f=2$ (subsection B). 
In Sec.~V, we give our conclusions.  
Some formulae and technical details used in our analysis 
 are summarized in Appendices A and B.

\section{Classical structure of ${\cal N}=2$ $SU(2)\times U(1)$ gauge theory}

In this section, we briefly discuss the classical structure of our theory. 
The complete analysis of the classical potential was 
 originally addressed in Ref. \cite{fayet}. 

We describe the classical Lagrangian in terms of ${\cal N}=1$ superfields: 
 adjoint chiral superfield $A_i$, superfield strength $W_i$ 
 and vector superfield $V_i$ in the vectormultiplet 
 ($i=1,2$ denote the index of 
 the $U(1)$ and the $SU(2)$ gauge symmetries, respectively),  
 and two chiral superfields $Q^i_\alpha$ and $\tilde{Q}_i^\alpha$ 
 in the hypermultiplet ($i=1,...,N_f$ is the flavor index, 
 and $\alpha=1,2$ is the $SU(2)$ color index). 
The classical Lagrangian is given by 
\begin{eqnarray}
{\cal L}&=&{\cal L}_{HM}
           +{\cal L}_{VM}
           +{\cal L}_{FI} \; ,  \label{eq:lag}
\end{eqnarray}
\begin{eqnarray}
{\cal L}_{HM}
   &=& \int d^4\theta  \left( 
       Q_i^\dagger e^{2V_2+2V_1} Q^i 
       + \tilde{Q}_i e^{-2V_2-2V_1} \tilde{Q}^{\dagger i}
          \right)  \nonumber \\  
   &+& \sqrt{2} \left( \int d^2 \theta 
           \tilde{Q}_i \left( A_2+A_1 \right) Q^i + h.c. \right) \; , \\
{\cal L}_{VM} 
        &=&\frac{1}{2\pi} \mbox{Im}\left[ \mbox{tr}
           \left\{\tau_{22}
           \left( \int d^4\theta A_2^\dagger e^{2 V_2} A_2 e^{-2 V_2}
           +\frac{1}{2}\int d^2 \theta  W_2^2 
 \right)\right\}\right] \nonumber \\
  &+& \frac{1}{4\pi}{\rm Im}\left[\tau_{11} \left(\int d^4\theta 
           A_1^\dagger A_1+\frac{1}{2} \int d^2\theta 
           W_1^2 \right)\right] \; , \\
{\cal L}_{FI} 
        &=&\int d^4\theta \xi V_1  \; ,      \label{eq:FI}
\end{eqnarray}
where $\tau_{22}=i\frac{4\pi}{g^2}+\frac{\theta}{2\pi}$ 
 and $\tau_{11}=i\frac{4\pi}{e^2}$ are the gauge couplings 
 of the $SU(2)$ and the $U(1)$ gauge interactions, respectively.
Here we take the notation, 
 $T(R) \delta^{ab}$=tr($T^a T^b)=\frac{1}{2}\delta^{ab}$. 
The same $U(1)$ charge of the hypermultiplets is normalized to be one.
The last term in Eq.~(\ref{eq:lag}) is the FI term 
 with the coefficient $\xi$ of mass dimension two. 

From the above Lagrangian, the classical potential is read off as
\begin{eqnarray} 
 V &=&\frac{1}{g^2}\mbox{\rm tr}[A_2, A_2^\dagger]^2 
       +\frac{g^2}{2} 
 ( q_i^{\dagger} T^a q^i- \tilde{q}_i T^a  \tilde{q}^{\dagger i} )^2  
 \nonumber \\
    &+& q_i^\dagger [A_2, A_2^\dagger] q^i  
     - \tilde{q}_i [A_2,A_2^\dagger] \tilde{q}^{\dagger i} 
    +2 g^2 |\tilde{q}_i T^a q^i |^2  \nonumber \\
   &+&\frac{e^2}{2} \left( \xi+ q_i^\dagger q^i- \tilde{q}_i 
       \tilde{q}^{\dagger i} \right)^2 
     +2 e^2 |\tilde{q}_i q^i|^2 \nonumber \\
   &+& 2 \left( q_i^\dagger |A_2+A_1|^2 q^i 
    +\tilde{q}_i |A_2+A_1|^2 \tilde{q}^{\dagger i} \right) \; ,
\end{eqnarray}
where $A_2$, $A_1$, $q^i$ and $\tilde{q}_i$ are scalar components 
 of the corresponding chiral superfields, respectively. 
The potential minimum is obtained by solving 
 the stationary conditions with respect to these scalar components. 
There are some solutions, and one example is given by
\begin{eqnarray}
& &  q^i_\alpha  = 0, \; \; 
    \tilde{q}_1^\alpha = \delta^\alpha_1 
     \left( \frac{e^2}{\frac{1}{4}g^2+e^2} \xi \right)^{\frac{1}{2}}, \; \; 
    \tilde{q}_j^\alpha = 0 \; \; (j \neq 1) ,  \nonumber \\   
& &  A_2 + A_1 = \left(\begin{array}{cc}
        \frac{a_2}{2} & 0 \\  0 & -\frac{a_2}{2}  \end{array}\right) 
            + \left(\begin{array}{cc} a_1 & 0 \\ 0 & a_1 \end{array}\right) 
            =  \left(\begin{array}{cc} 0 & 0 \\
                     0 & z  \end{array}\right),
        \label{eq:mix}
\end{eqnarray}
where $a_1$ and $a_2$ are complex parameters, 
 and $z$ is arbitrary constant.  
In this example, the gauge symmetry $SU(2)\times U(1)$ is broken to $U(1)$. 
The potential energy is given by 
\begin{eqnarray}
V=\frac{\xi^2}{2} \frac{e^2 g^2}{4e^2+g^2}.
\end{eqnarray}

Note that the classical potential has the pseudo flat direction  
 parameterized by $a_1$ or $a_2$ 
 with the condition $a_1+\frac{1}{2}a_2=0$. 
We expect that this direction is lifted up, 
 once quantum corrections are take into account, 
 and the true non-degenerate vacuum is selected out 
 after the effective potential is analyzed. 
This naive expectation seems natural, 
 if we notice that the above potential energy is 
 described by the bare gauge couplings, 
 which should be replaced by the effective one 
 (non-trivial functions of moduli parameters) 
 in the effective theory.

\section{Quantum structure of ${\cal N}=2~SU(2)\times U(1)$ gauge theory}

\subsection{Effective Action}

In this subsection, we describe the low energy Wilsonian effective 
 Lagrangian of our theory. 
If we could completely integrate the action to zero momentum, 
 the exact effective Lagrangian ${\cal L}_{EXACT}$ could be obtained, 
 which is described by light fields, the dynamical scale 
 and the coefficient of the FI term $\xi$. 
However, this is highly non-trivial and very difficult task. 
In the following discussion, suppose that the coefficient $\xi$, 
 the order parameter of SUSY breaking, 
 is much smaller than the dynamical scale of the $SU(2)$ gauge interaction. 
Then we consider the effective action up to the leading order of $\xi$. 
The exact effective Lagrangian, if it could be obtained, 
 can be expanded with respect to the parameter $\xi$ as 
\begin{eqnarray}
{\cal L}_{EXACT}
        ={\cal L}_{SUSY}
         +\xi {\cal L}_1 + {\cal O}(\xi^2).
         \label{eq:eff}
\end{eqnarray}
Here, the first term ${\cal L}_{SUSY}$ 
 is the exact effective Lagrangian containing full SUSY quantum corrections. 
The second term is the leading term of $\xi$, 
 and nothing but the FI term at tree level.
\footnote{
 Considering all the symmetries of our theory, 
 we find that the FI term is tree-level exact \cite{arai}. } 
Analyzing the effective Lagrangian up to the leading order of $\xi$,
 we obtain the effective potential of the order of $\xi^2$. 
The coefficient of $\xi^2$ in the effective potential includes 
 full SUSY quantum corrections. 
Therefore, in our aim, what we need to analyze the effective potential 
 is nothing but the effective Lagrangian ${\cal L}_{SUSY}$. 

Except the FI term, the classical $SU(2)\times U(1)$ gauge theory 
 has moduli space, which is parameterized by $a_2$ and $a_1$. 
On this moduli space except the origin, the gauge symmetry is broken to 
 $U(1)_c\times U(1)$. 
Here $U(1)_c$ denotes the gauge symmetry in the Coulomb phase 
 originated from the $SU(2)$ gauge symmetry. 
Before discussing the effective action of this theory, 
 we should make it clear how to treat the $U(1)$ gauge interaction part. 
In the following analysis, this part is, as usual, discussed 
 as a cut-off theory.
\footnote{
 There is a possibility that 
 non-trivial fixed point and the strong coupling phase 
 exist in QED \cite{miransky}. 
 This problem is very difficult, and is out of our scope 
 (see also Ref.~\cite{arai} for related discussions).} 
Thus, the Landau pole $\Lambda_L$ is inevitably introduced 
 in our effective theory, and the defined region 
 of the moduli parameter $a_1$ is constrained 
 within the region $|a_1|< \Lambda_L$. 
According to this constraint, the defined region for moduli parameter 
 $a_2$ is found to be also constrained in the same region, 
 since two moduli parameters are related with each other 
 through the hypermultiplets. 
We take the scale of $\Lambda_L$ to be much larger than 
 the dynamical scale of the $SU(2)$ gauge interaction $\Lambda_{N_f}$, 
 so that the $U(1)$ gauge interaction is always weak 
 in the defined region of moduli space. 
Note that, in our framework, we implicitly assume that 
 the $U(1)$ gauge interaction have no effect on the $SU(2)$ gauge dynamics.
This assumption is justified in the following discussion about 
 the monodromy transformation (see Eq.~(\ref{eq:mono})). 

We first discuss the general formulae for the effective Lagrangian 
 ${\cal L}_{SUSY}$, which 
 consists of two parts described by light vectormultiplets 
 and hypermultiplets, 
 ${\cal L}_{SUSY}={\cal L}_{VM}+{\cal L}_{HM}$. 
The vectormultiplet part ${\cal L}_{VM}$, 
 which is consistent with ${\cal N}=2$ SUSY and all the symmetries 
 in our theory, is given by  
\begin{eqnarray}
{\cal L}_{VM}
        =\frac{1}{4\pi}
           \mbox{\rm Im}\left\{
           \sum_{i,j=1}^{2}\left(\int d^4\theta
           \frac{\partial F}{\partial A_i} A_i^\dagger
         + \int d^2\theta \frac{1}{2} 
           \tau_{ij} W_i W_j \right) \right\} ,
\end{eqnarray}
where $F(A_2,A_1,\Lambda_{N_f},\Lambda_L)$ 
 is the prepotential, which is the function of moduli parameters  
 $a_2$, $a_1$, the dynamical scale $\Lambda_{N_f}$, 
 and the Landau pole $\Lambda_L$. 
The effective coupling $\tau_{ij}$ is defined as 
\begin{eqnarray}
\tau_{ij}=
   \frac{\partial^2 F}{\partial a_i \partial a_j} \; \; (i,j=1,2) .  
   \label{eq:coupling}
\end{eqnarray}
The part ${\cal L}_{HM}$ is 
 described by a light hypermultiplet 
 with appropriate quantum number $(n_e,n_m)_n$, 
 where $n_e$ is electric charge, $n_m$ is magnetic charge, 
 and $n$ is the $U(1)$ charge.  
This part should be added to the effective Lagrangian 
 around a singular point on moduli space, 
 since the hypermultiplet is expected to be light there  
 and enjoys correct degrees of freedom in the effective theory. 
The explicit description is given by 
\begin{eqnarray}
{\cal L}_{HM}
 &=& \int d^4 \theta \left( 
      M^\dagger e^{2 n_m V_{2D}+2 n_e V_2 + 2 n V_1}M
      +\tilde{M} e^{-2 n_m V_{2D}-2 n_eV_2-2 n V_1} \tilde{M}^{\dagger} 
              \right)  \\ \nonumber
 &+& \sqrt{2} \left( \int d^2\theta \tilde{M} (n_m A_{2D}+n_e A_2+n A_1) M 
               +h.c. \right) \; , 
\end{eqnarray}
where $M$ and $\tilde{M}$ denote light quark or light dyon hypermultiplet, 
 that is, the light BPS state, 
 and $V_{2D}$ is the dual gauge field of $U(1)_c$. 

In order to obtain an explicit description of the effective Lagrangian, 
 let us consider the monodromy transformation of our theory. 
Suppose that moduli space is parameterized by the vectormultiplet scalars 
 $a_2$, $a_1$ and their duals $a_{2D}$, $a_{1D}$ 
 which are defined as $a_{iD}=\partial F/ \partial a_i$ ($i=1,2$). 
These variables are transformed into their linear combinations 
 by the monodromy transformation. 
In our case, the monodromy transformation is subgroup 
 of $Sp(4,\mbox{\bf R})$, which leaves the effective Lagrangian invariant, 
 and the general formula is found to be \cite{marino3} 
\begin{eqnarray}
 \left(\begin{array}{c}
       a_{2D} \\  a_{2}  \\  a_{1D} \\  a_1  \end{array}  \right)
 \rightarrow 
 \left(\begin{array}{c}
       \alpha a_{2D}+\beta a_2+p a_1   \\
       \gamma a_{2D}+\delta a_2+q a_1  \\
       a_{1D}+p(\gamma a_{2D}+\delta a_2)
        -q(\alpha a_{2D}+\beta a_2)-pqa_1\\ a_1 
\end{array}  \right),  
  \label{eq:mono}
\end{eqnarray}
where 
$
\left(\begin{array}{cc}
\alpha & \beta  \\ \gamma & \delta \end{array} \right) 
 \in SL(2,\mbox{\bf Z})
$
 and $p,q \in\mbox{\bf Q}$. 
Note that this monodromy transformation for the combination 
 $(a_{2D},a_{2},a_1)$ is exactly the same as that 
 for SQCD with massive quark hypermultiplets, 
 if we regard $a_1$ as the same mass of the hypermultiplets 
 such that $m= \sqrt{2} a_1$.  
This fact means that the $U(1)$ gauge interaction part plays 
 the only role as the mass term for the $SU(2)$ gauge dynamics. 
This observation is consistent with our assumption. 
On the other hand, the $SU(2)$ dynamics plays an important role 
 for the $U(1)$ gauge interaction part, 
 as can be seen in the transformation law of $a_{1D}$. 
This monodromy transformation is also used to derive dual variables 
 associated with the BPS states. 
As a result, the prepotential of our theory turns out to be 
 essentially the same as the result in \cite{s-w2} 
 with understanding the relation $A_1 = m/\sqrt{2}$, 
\begin{eqnarray}
F(A_2,A_1,\Lambda_{N_f},\Lambda_L)
   =F_{SU(2)}^{(SW)}(A_2, m,\Lambda_{N_f})
    \Bigg{|}_{A_1=\frac{m}{\sqrt{2}}}+C A_1^2,
    \label{eq:pre}
\end{eqnarray}
where the first term is the prepotential of ${\cal N}=2$ SQCD 
 with hypermultiplets having the same mass $m$, 
 and $C$ is free parameter. 
The freedom of the parameter $C$ is used to determine 
 the scale of the Landau pole 
 relative to the scale of the $SU(2)$ dynamics.

\subsection{Effective Potential}

The effective potential can be read off from the above Lagrangian 
 with the FI term such that
\footnote{
 We suppose that the potential is described by the adequate variables
 associated with the light BPS states. 
 For instance, the variable $a_2$ is understood implicitly as $-a_{2D}$, 
 when we consider the effective potential for the monopole.} 
%
\begin{eqnarray}
V &=& b_{11}|F_1|^2  + b_{12}(F_1 F_2^\dagger + F_1^\dagger F_2) 
     + b_{22}|F_2|^2 
     +\frac{1}{2}b_{11}D_1^2 +b_{12}D_1D_2+\frac{1}{2}b_{22}D_2^2  \nonumber \\
 &+& (D_2+nD_1)
     (|M|^2- |\tilde{M}|^2)
     +|F_M|^2+|F_{\tilde{M}}|^2 \\ 
 &+& \sqrt{2}(F_2 M \tilde{M}
     + a_2  M F_{\tilde{M}} + a_2 \tilde{M} 
       F_M + h.c.) \nonumber \\
 &+& \sqrt{2}(n F_1 M \tilde{M}+ n a_1 M F_{\tilde{M}}
     + n a_1 \tilde{M} F_{M} + h.c.)
     +\xi D_1 \; ,  \nonumber
\end{eqnarray}
where $F_I (I=1, 2, M, \tilde{M})$ denotes the auxiliary field  
 of the corresponding chiral multiplet, 
 $D_J (J=1, 2)$ denotes the auxiliary field  
 of the corresponding vectormultiplet $V_J$,
 and the effective gauge coupling is defined as 
 $b_{ij}=(1/4 \pi) \mbox{Im} \tau_{ij}$. 
Eliminating these auxiliary fields by using their equations of motion, 
\begin{eqnarray}
D_1&=&\frac{1}{\det b}
      \left\{(b_{12}-nb_{22})
      (|M|^2-|\tilde{M}|^2)-\xi b_{22}\right\},
      \nonumber \\
D_{2}&=&\frac{1}{\det b}
      \left\{-(b_{11}-nb_{12})(|M|^2-|\tilde{M}|^2)
      +\xi b_{12}\right\}, \nonumber \\
F_1&=&\frac{\sqrt{2}}{\det b}(b_{12}-nb_{22})
      ( M \tilde{M})^\dagger, \\ 
F_2&=&\frac{\sqrt{2}}{\det b}
      (nb_{12}-b_{11}) (M \tilde{M})^\dagger, 
      \nonumber \\
F_{M} &=& -\sqrt{2}(a_2^\dagger
      + n a_1^\dagger)\tilde{M}^{\dagger},  \nonumber \\
F_{\tilde{M}}
   &=& -\sqrt{2}(a_2^\dagger
      + n a_1^\dagger) M^\dagger, \nonumber
\end{eqnarray}
where $\det b= b_{22} b_{11} -b_{12}^2$, 
 we obtain 
\begin{eqnarray}
V &=& \frac{b_{22}}{2\det b}\xi^2
      +S(a_2,a_1)\left\{(|M|^2 - |\tilde{M}|^2)^2
                 +4|M \tilde{M}|^2\right\} \nonumber \\ 
 &+&  2 T(a_2,a_1)(|M|^2 + |\tilde{M}|^2)
      -U(a_2,a_1) (|M|^2-|\tilde{M}|^2),
\end{eqnarray}
where $S$, $T$ and $U$ are defined as 
\begin{eqnarray}
S(a_2,a_1)&=&\frac{1}{2b_{22}}
            +\frac{(b_{12}-nb_{22})^2}
                  {2b_{22}\det b}, \\
T(a_2,a_1)&=&|a_2+na_1|^2, \\
U(a_2,a_1)&=&\frac{b_{12}-nb_{22}}{\det b}\xi.
\end{eqnarray}
The stationary conditions with respect to the hypermultiplets, 
\begin{eqnarray}
\frac{\partial V}{\partial M^\dagger}
 &=& M \left\{2 S (|M|^2-|\tilde{M}|^2) + 2T - U \right\}
   +4 S \tilde{M}^{\dagger} (M \tilde{M})=0,  
 \label{eq:mini1} \\
\frac{\partial V}{\partial \tilde{M}^{ \dagger}} 
 &=& \tilde{M} \left\{-2 S (|M|^2-|\tilde{M}|^2) + 2T + U  \right\}
   + 4 S M^{\dagger} (M \tilde{M})=0,  
 \label{eq:mini2}
\end{eqnarray}
 lead to three solutions as follows: 
\begin{eqnarray}
1. \;  & & M = \tilde{M} =0;  \; \; 
  V=\frac{b_{22}}{2\det b}\xi^2,  \label{eq:sol1} \\
2. \; & & |M|^2=-\frac{2T-U}{2S}, \; \tilde{M}=0; \; \; 
  V=\frac{b_{22}}{2\det b}\xi^2-S |M|^4.
       \label{eq:sol2} \\
3. \; & & M=0, \; |\tilde{M}|^2 = -\frac{2T+U}{2S};  \; \; 
  V=\frac{b_{22}}{2\det b}\xi^2 -S |\tilde{M}|^4 .  \label{eq:sol3}
\end{eqnarray}
The solution Eq.~(\ref{eq:sol2}) or Eq.~(\ref{eq:sol3}), 
 in which the light hypermultiplet acquires 
 the vacuum expectation value, 
 is energetically favored, because of $\det b >0$ and $S(a_2,a_1)>0$.  
Since the hypermultiplet appears in the theory as the light BPS state  
 around the singular point on moduli space, 
 the potential minimum is expected to emerge there. 
On the other hand, the solution Eq.~(\ref{eq:sol1}) describes 
 the potential energy away from the singular points, 
 which smoothly connects with the solution 
 Eq.~(\ref{eq:sol2}) or Eq.~(\ref{eq:sol3}). 

\subsection{Periods and Effective Couplings}

It was shown that the effective potential is described by the periods 
 $a_{2D}$, $a_{2}$ and the effective gauge coupling $b_{ij}$. 
In this subsection, we derive the periods and the effective gauge couplings 
 in order to give an explicit description of the effective potential. 
As already discussed, the periods are the same as that of SQCD, 
 which were derived in both cases 
 with the massless \cite{klemm,yang,ferrari1,isidro} and 
 the massive \cite{marino3,masuda,bran,ferrari2} hypermultiplets. 
There are some different descriptions of the periods 
 with such as the Weierstrass functions \cite{marino3},  
 the hypergeometric functions \cite{masuda}, 
 the modular functions \cite{bran} and 
 the elliptic integrals \cite{ferrari2}. 
In our analysis, we use the integral representation. 
On the other hand, the effective coupling $\tau_{ij}$ 
 is described in terms of the Weierstrass functions. 

\begin{center}
\begin{figure}[p]
\centerline{\psfig{file=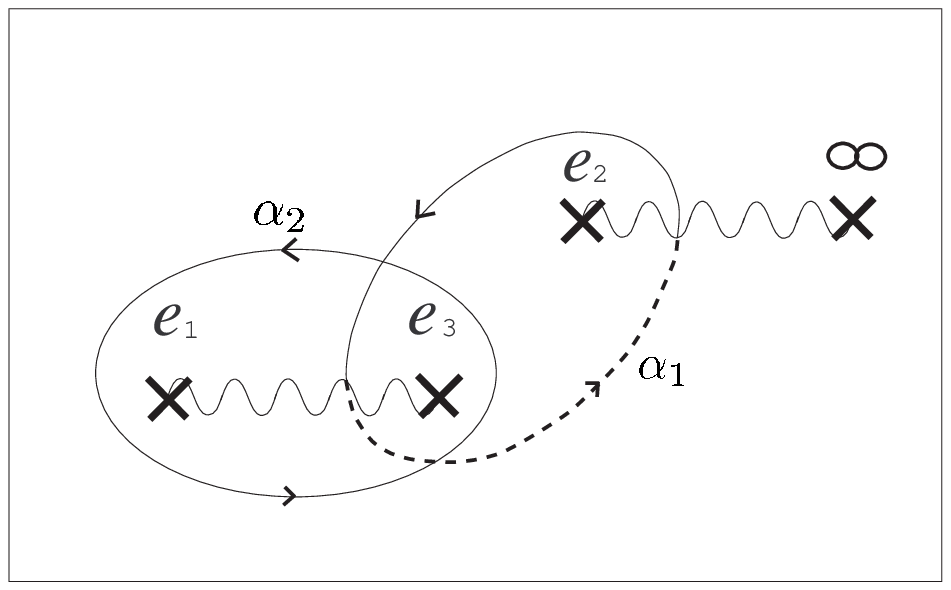,width=11cm}}
\caption{ The contours $\alpha_1$ and $\alpha_2$. }
\label{contour1}
\end{figure}
\end{center}

We first review how to obtain the periods $a_{2D}$ and  $a_2$.
The elliptic curves of ${\cal N}=2$ SQCD with hypermultiplets 
 having the same mass $m$ were found to be \cite{s-w2} 
\begin{eqnarray}
y^2=x^2(x-u)+P_{N_f}(x,u,m, \Lambda_{N_f}), 
 \label{eq:curve}
\end{eqnarray}
where the polynomials $P_{N_f}(N_f=1,2)$ are given by
\begin{eqnarray}
P_1&=&\frac{\Lambda_1^3}{4}mx
     -\frac{\Lambda_1^6}{64},\\
P_2&=&-\frac{\Lambda_2^4}{64}(x-u)
       +\frac{\Lambda_2^2}{4}
       m^2x-\frac{\Lambda_2^4}{32}m^2.
\end{eqnarray}
In this case, the mass formula of the BPS state 
 with the quantum number $(n_e, n_m)_n$ 
 is given by $M_{BPS}=\sqrt{2}| n_m a_{2D}+n_e a_2+n m/\sqrt{2}|$. 
If $\lambda$ is a meromorphic differential on the 
curve Eq.~(\ref{eq:curve}) such that
\begin{eqnarray}
\frac{\partial \lambda}{\partial u}
       =\frac{\sqrt{2}}{8\pi}\frac{dx}{y},
\end{eqnarray}
the periods are given by the contour integrals
\begin{eqnarray}
  a_{2D}=\oint_{\alpha_1}\lambda, \; 
  a_{2}=\oint_{\alpha_2}\lambda,
 \label{eq:period}
\end{eqnarray}
where the cycles $\alpha_1$ and $\alpha_2$ are defined 
 so as to encircle $e_2$ and $e_3$, and $e_1$ and $e_3$, respectively
 (see. Fig.~\ref{contour1}). 
Meromorphic differentials are given by  
\begin{eqnarray}
\lambda_{SW}^{(N_f=1)}
     &=&-\frac{\sqrt{2}}{4\pi}\frac{ydx}{x^2} 
      =\frac{\sqrt{2}}{4\pi} \left[ 
     -\frac{dx}{2y} \left(3x-2u+\frac{m\Lambda_1^3}{4x}\right)
      + dx\frac{d}{dx}\left(\frac{x}{y}\right) \right],
     \label{eq:f1}  
  \\
\lambda_{SW}^{(N_f=2)}
     &=&-\frac{\sqrt{2}}{4\pi}
        \frac{ydx}{x^2-\frac{\Lambda^4}{64}} 
      =-\frac{\sqrt{2}}{4\pi}\frac{dx}{y}
        \left[ x-u  +\frac{m^2\Lambda_2^2} 
         {4\left(x+\frac{\Lambda^2}{8} \right)}\right].
     \label{eq:f2}
\end{eqnarray}
Each differential have the single pole at $x=0$ for $N_f=1$ 
 and $x=-\frac{\Lambda_2}{8}$ for $N_f=2$. 
For both cases, the residue is given by 
\begin{eqnarray}
\mbox{\rm Res}\lambda_{SW}^{(N_f)}
   =\frac{1}{2\pi i}(-1)\frac{m}{\sqrt{2}}.
\end{eqnarray} 

We calculate the periods by using the Weierstrass normal form 
 for later convenience. 
In this form, the algebraic curve is rewritten 
 by new variables $x=4X+\frac{u}{3}$ and $y=4Y$, such that 
\begin{eqnarray}
Y^2 = 4X^3-g_2^{(N_f)}X-g_3^{(N_f)}
   &=&4(X-e_1)(X-e_2)(X-e_3),\\ \nonumber
\sum_{i=1}^3e_i&=&0,
\end{eqnarray}
where $g_2^{(N_f)}$ and $g_3^{(N_f)}$ are explicitly written by 
\begin{eqnarray}
g_2^{(1)}&=&\frac{1}{4}\left(\frac{u^2}{3}
                   -\frac{m\Lambda_1^3}{3}\right),\\
g_3^{(1)}&=&\frac{1}{16}\left(-\frac{mu\Lambda_1^3}{12}
                   +\frac{\Lambda_1^6}{64}+\frac{2u^3}{27}\right),\\
g_2^{(2)}&=&\frac{1}{16}\left(\frac{4}{3}u^2
                   +\frac{\Lambda_2^4}{16}-m^2 
                    \Lambda_2^2\right),\\
g_3^{(2)}&=&\frac{1}{16}\left(
                   \frac{m^2 \Lambda_2^4}{32}
                   -\frac{u}{12}m^2\Lambda_2^2
                   -\frac{u\Lambda_2^4}{96}
                   +\frac{2u^3}{27}\right).
\end{eqnarray}
Converting the Seiberg-Witten differentials of 
 Eqs.(\ref{eq:f1}) and (\ref{eq:f2}) 
 into the Weierstrass normal form 
 and substituting them into Eq.(\ref{eq:period}), 
 we obtain the integral representations of the periods as follows 
 ($a_{2D}$ and $a_{2}$ are denoted by $a_{21}$ and $a_{22}$, respectively):
\begin{eqnarray}
 a_{2i}^{(N_f=1)}&=&-\frac{\sqrt{2}}{4\pi}
    \left(-uI_1^{(i)}+12I_2^{(i)}
    +\frac{m\Lambda_1^3}{16}I_3^{(i)}(c)\right),
    \\ 
 a_{2i}^{(N_f=2)}&=&-\frac{\sqrt{2}}{4\pi}
    \left(-\frac{4}{3}uI_1^{(i)}+8I_2^{(i)}
    +\frac{m^2\Lambda_2^2}{8}
    I_3^{(i)}
    \left(c \right)\right), \end{eqnarray}
where $c$ is the pole of the differentials, 
 and $c=-\frac{u}{12}$ for $N_f=1$ and 
 $c=-\frac{u}{12}-\frac{\Lambda_2^2}{32}$ for $N_f=2$.  
Integrals $I_1^{(i)},I_2^{(i)}$ and $I_3^{(i)}$ are defined as 
\begin{eqnarray}
I_1^{(i)}=\frac{1}{2}\oint_{\alpha_i}\frac{dX}{Y}, \; \; 
I_2^{(i)}=\frac{1}{2}\oint_{\alpha_i}\frac{XdX}{Y}, \; \; 
I_3^{(i)}(c)=\frac{1}{2}
            \oint_{\alpha_i}\frac{dX}{Y(X-c)}.
\end{eqnarray}
The roots $e_i$ of the polynomial defining the cubic 
 are chosen so as to lead to the correct asymptotic behavior 
 for large $|u|$, 
\begin{eqnarray}
a_{2D}^{(N_f)}(u) 
 \sim i\frac{4-N_f}{2\pi}\sqrt{2u} \log\frac{u}{\Lambda_{N_f}^2}, \; \; 
a_2^{(N_f)}(u)\sim\frac{\sqrt{2u}}{2}.
\end{eqnarray}
A correct choice is the following: 
\begin{eqnarray}
N_f=1~\mbox{\rm case:}& & \nonumber \\ 
e_1&=&\frac{1}{48}
      \frac{-24 \Lambda_1^3m+32u^2+2^{1/3}H(u,m,\Lambda_1)^{2/3}}
           {H(u,m,\Lambda_1)^{1/3}},\nonumber \\ 
e_2&=&\frac{1}{96}
      \frac{8(1-3i)(3\Lambda_1^2m-4u^2)-2^{1/3}i(-i+\sqrt{3})
                    H(u,m,\Lambda_1)^{2/3}}
           {2^{2/3}H(u,m,\Lambda_1)^{1/3}},\nonumber \\ 
e_3&=&\frac{1}{96}
      \frac{8(1+3i)(3\Lambda_1^2m-4u^2)+2^{1/3}i(i+\sqrt{3})
                    H(u,m,\Lambda_1)^{2/3}}
           {2^{2/3}H(u,m,\Lambda_1)^{1/3}}, \\ 
H(u,m,\Lambda_1)&=&27\Lambda_1^6+128u^3 \nonumber \\
   & &+3\Lambda_1^3
        (-48mu+\sqrt{3}\sqrt{27\Lambda_1^6-256(m^2-u)u^2}
        +32\Lambda_1^3(8m^3-9mu)), \nonumber \\ 
N_f=2~\mbox{\rm case:}& & \nonumber \\ 
e_1 &=&\frac{u}{24}-\frac{\Lambda_2^2}{64}
     -\frac{1}{8}\sqrt{u+\frac{\Lambda_2^2}{8}
                  +\Lambda_2 m}
                 \sqrt{u+\frac{\Lambda_2^2}{8}
                  -\Lambda_2 m},
                 \nonumber \\
e_2 &=&\frac{u}{24}-\frac{\Lambda_2^2}{64}
     +\frac{1}{8}\sqrt{u+\frac{\Lambda_2^2}{8} + \Lambda_2 m}
                 \sqrt{u+\frac{\Lambda_2^2}{8} - \Lambda_2 m},
                 \label{eq:root}\\
e_3 &=&-\frac{u}{12}+\frac{\Lambda_2^2}{32}. 
                 \nonumber
\end{eqnarray}
Fixing the contours of the cycles relative to the positions 
 of the poles, which is equivalent to fix the $U(1)$ charges 
 (baryon numbers) for the BPS states, 
 the final formulae are given by 
\begin{eqnarray}
a_{2i}^{(N_f=1)}&=&-\frac{\sqrt{2}}{4\pi}
    \left(-uI_1^{(i)}+12I_2^{(i)}
    +\frac{m\Lambda_1^3}{16}I_3^{(i)}
     \left(-\frac{u}{12}\right)\right)
    -\frac{m}{\sqrt{2}}\delta_{i2}, 
    \label{eq:period1}
    \\ 
a_{2i}^{(N_f=2)}&=&-\frac{\sqrt{2}}{4\pi}
    \left(-\frac{4}{3}uI_1^{(i)}+8I_2^{(i)}
    +\frac{m^2\Lambda_2^2}{8}
    I_3^{(i)}
    \left(-\frac{u}{12}
    -\frac{\Lambda_2^2}{32}\right)\right)
    -\frac{m}{\sqrt{2}}\delta_{i2},
    \label{eq:period2} 
\end{eqnarray}
with the integral $I_i^{(1)} \; (i=1,2)$ explicitly given by
\begin{eqnarray}
I_1^{(1)}&=&\int_{e_2}^{e_3}\frac{dX}{Y}
          = \frac{iK(k^\prime)}{\sqrt{e_2-e_1}}, 
            \label{eq:formula1} \\
I_2^{(1)}&=&\int_{e_2}^{e_3}\frac{XdX}{Y}
          = \frac{ie_1}{\sqrt{e_2-e_1}}K(k^\prime)
            +i\sqrt{e_2-e_1}E(k^\prime), 
            \label{eq:formula2} \\
I_3^{(1)}&=&\int_{e_2}^{e_3}\frac{dX}{Y(X-c)}
             = \frac{-i}{(e_2-e_1)^{3/2}}
            \left\{
            \frac{1}{k+\tilde{c}}K(k^\prime)
            +\frac{4k}{1+k}
            \frac{1}{\tilde{c}^2-k^{2}}
            \Pi_1\left(\nu,\frac{1-k}{1+k}         
                \right)            
            \right\},
            \label{eq:formula3}
\end{eqnarray}
where $k^2 = \frac{e_3-e_1}{e_2-e_1}$, 
      $k^{\prime 2}=1-k^2=\frac{e_2-e_3}{e_2-e_1}$,  
      $\tilde{c}= \frac{c-e_1}{e_2-e_1}$, 
and $\nu=-\left(\frac{k+\tilde{c}}{k-\tilde{c}}\right)^2
          \left(\frac{1-k}{1+k}\right)^2$. 
The formulae for $I_i^{(2)}$ are obtained from $I_i^{(1)}$ 
 by exchanging the roots, $e_1$ and $e_2$. 
In Eqs.~(\ref{eq:formula1})-(\ref{eq:formula3}), 
 $K$, $E$, and $\Pi_1$ are the complete elliptic integrals \cite{higher} 
 given in Appendix A. 

\begin{center}
\begin{figure}[p]
\centerline{\psfig{file=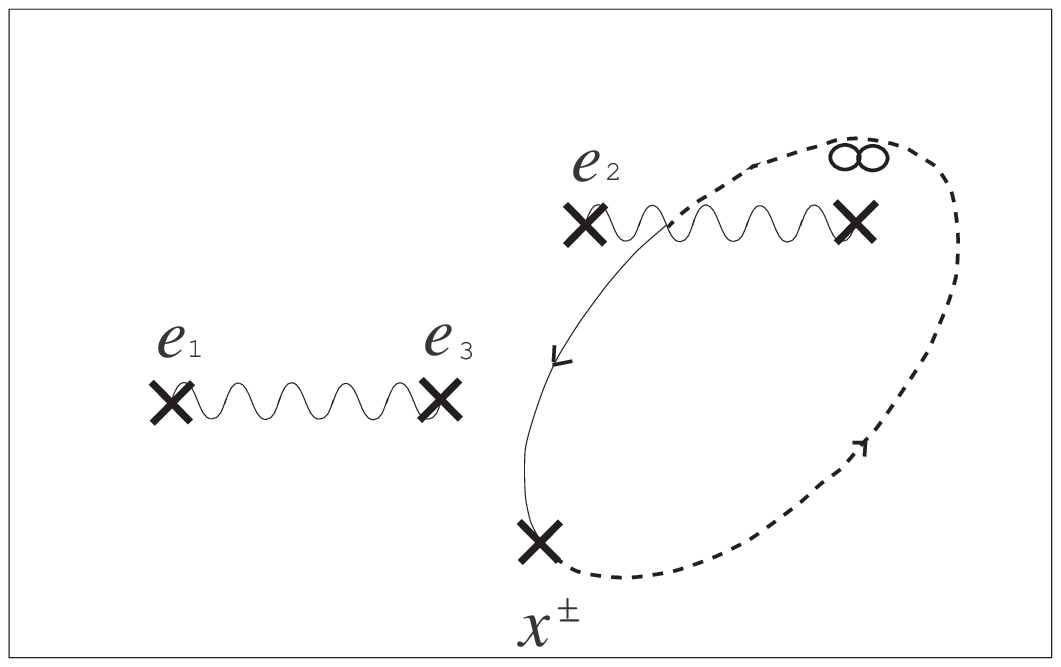,width=10cm}}
\caption{ The contour of the integral Eq.~(\ref{eq:bilinear}).}
\label{contour2}
\end{figure}
\end{center}

Next let us consider the effective coupling defined as 
 Eq.~(\ref{eq:coupling}). 
The effective couplings $\tau_{22}$ and $\tau_{12}(= \tau_{21})$ 
 are obtained by 
\begin{eqnarray}
\tau_{22}&=&\frac{\partial a_{2D}}{\partial a_2}=\frac{\omega_1}{\omega_2},\\
\tau_{12}&=&\frac{\partial a_{2D}}{\partial a_1}
          \Bigg{|}_{a_2}
         =\frac{\partial a_{2D}}{\partial a_1}
           \Bigg{|}_u
          -\tau_{22}
          \frac{\partial a_{2}}{\partial a_1}
           \Bigg{|}_u
         =-\frac{2z_0}{\omega_2},
         \label{eq:tau12}
\end{eqnarray}
where $\omega_i$ is the period of the Abelian differential, 
\begin{eqnarray}
\omega_i=\oint_{\alpha_i}\frac{dX}{Y}=2I_1^{(i)} \; ~(i=1,2) ,
\end{eqnarray}
and $z_0$ is defined as
\begin{eqnarray}
z_0=-\frac{1}{\sqrt{e_2-e_1}}F(\phi,k); \; \; 
 \sin^2\phi=\frac{e_2-e_1}{c-e_1} \; . 
\end{eqnarray}
Here $F(\phi,k)$ is the incomplete elliptic integral of 
 the first kind given in Appendix B. 
The effective coupling $\tau_{11}$ is described 
 in terms of the Weierstrass function. 
First consider the period $a_{1D}$ 
 by using the Riemann bilinear relation \cite{riemann}, 
\begin{eqnarray}
\oint_{\alpha_1}\phi\oint_{\alpha_2}\omega
     -\oint_{\alpha_1}\omega\oint_{\alpha_2}\phi
     =2\pi i\sum_{n=1}^{N_p}
      \mbox{\rm Res}_{x_n^+}\phi
      \int_{x_n^-}^{x_n^+}\omega, \label{eq:bilinear}
\end{eqnarray}
where $\phi$ and $\omega$ are meromorphic and 
 holomorphic differentials, respectively, 
 $N_p$ is the number of poles ($N_p=1$ in our case), 
 and $x_n^{\pm}$ are poles of $\phi$ on the positive and 
 negative Riemann sheets. 
Substituting 
 $ \phi={\partial \lambda_{SW}^{(N_f)}}/{\partial a_1}$ 
 and $ \omega={\partial \lambda_{SW}^{(N_f)}}/{\partial a_2}$
 into Eq.~(\ref{eq:bilinear}), we obtain 
 (see Fig.~\ref{contour2} for the definition of the contour) 
\begin{eqnarray}
a_{1D}^{(N_f)}=-\sum_{n=1}^{N_p} \int_{x_n^-}^{x_n^+}
     \lambda_{SW} + \tilde{C} \; , \label{eq:mass}
\end{eqnarray}
where $\tilde{C}$ is a constant independent of $a_2$. 
The effective coupling $\tau_{11}$ is obtained 
 by differentiating Eq.~(\ref{eq:mass}) with respect to $a_1$  
 with $a_2$ fixed. 
This integral can be evaluated by the uniformization method 
 discussed in Appendix B. 
After regularizing the integral by using the freedom 
 of the constant $\tilde{C}$, 
 we finally obtain (see also Appendix B for details) 
\begin{eqnarray}
\tau_{11}=-\frac{1}{\pi i}\left[\log\sigma(2z_0)
           +\frac{4z_0^2}{\omega_2}
           I_1^{(2)}\right] + C ,  
         \label{eq:tau11}
\end{eqnarray}
where $\sigma$ is the Weierstrass sigma function, 
 and $C$ is the constant in Eq.~(\ref{eq:pre}). 

\begin{center}
\begin{figure}[p]
\centerline{\psfig{file=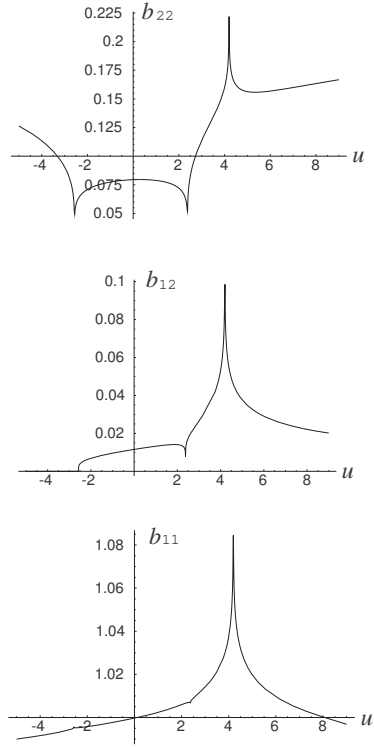,width=8cm}}
\caption{ The effective gauge couplings $b_{ij}$ 
 for $a_1=\sqrt{2}$ along real $u$-axis.}
\label{couplings}
\end{figure}
\end{center}

Note that, since the gauge coupling $b_{11}$ is found to be 
 a monotonically decreasing function of large $|a_1|$ 
 with fixed $u$, and vice versa  
 (see, for example, Fig.~\ref{couplings} in the case of fixed $a_1$), 
 the scale of the Landau pole is defined as $|a_1|=\Lambda_L$ 
 at which $b_{11}=0$. 
The large $\Lambda_L$ required by our assumption 
 is realized by taking an appropriate value for $C$. 
In the following analysis, we fix $C =4 \pi i $, 
 which corresponds to $\Lambda_L \sim 10^{17-18}$ 
 for fixed $\Lambda_{N_f} \sim O(1)$ 

In $N_f=1$ case, we plot the effective couplings $b_{ij}$ 
 along real $u$-axis in Fig.~\ref{couplings}. 
Here, the dynamical scale is normalized as $\Lambda_1= (256/27)^{1/6}$, 
 and the parameter $a_1 =\sqrt{2}$ is fixed. 
As expected, in the figure of $b_{22}$, 
 there are three singular points: 
 the dyon point ($u\sim -2.6$), the monopole point ($u\sim 2.4$) 
 and the quark singular point ($u\sim 4.2$). 
While the existence of the quark singular point is understood 
 based on the perturbative discussion, 
 the appearance of the dyon and the monopole singular points 
 is the result from the $SU(2)$ dynamics. 
Note that, in addition to the quark singular point, 
 there appear two singular points in the figures of $b_{12}$ and $b_{11}$. 
This result means that the $SU(2)$ dynamics plays an important role 
 for the $U(1)$ gauge interaction part in the infrared region 
 on the moduli space, as pointed out in the subsection A. 

\section{Potential Analysis}

Based on the results given by the previous sections, 
let us now investigate the vacuum structure of our theory. 
Since the effective potential is the function of 
 two complex moduli parameters $u$ and $a_1$, 
 it is a very complicated problem to figure out 
 behaviors of the effective potential 
 in the whole parameter space. 
However, note that, for our aim it is enough 
 to evaluate the potential energy just around the singular points, 
 since these points are energetically favored 
 (see Eqs.~(\ref{eq:sol1})-(\ref{eq:sol3})). 
The singular points on the moduli space parameterized by $u$ 
 flow according to the variation of $a_1$. 
In the following discussion, we evaluate the effective potential 
 along the flow of the singular points, 
 and examine which point is energetically favored 
 on the line of the flow.  

\subsection{Vacuum Structure in $N_f=1$ case} 

\begin{center}
\begin{figure}[p]
\centerline{\psfig{file=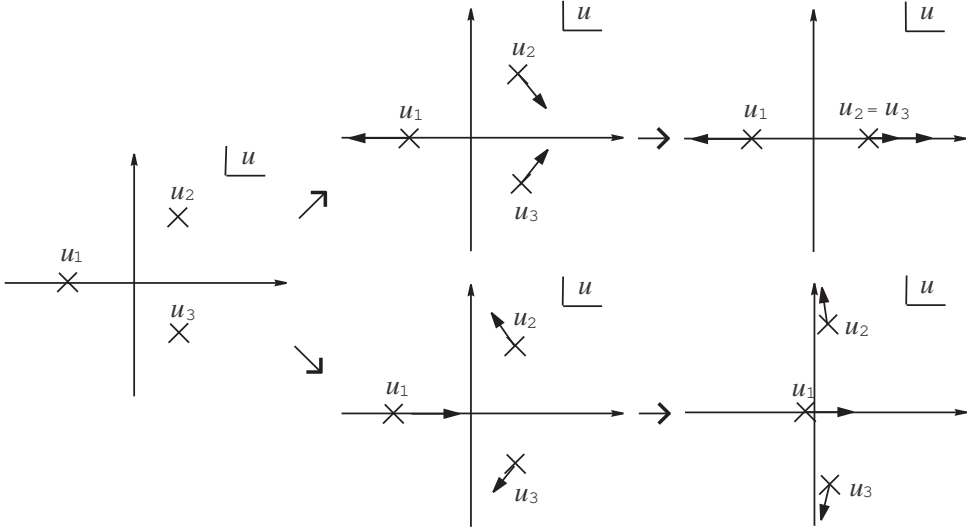,width=16cm}}
\caption{ The flow of the singular points on $u$-plane 
  for fixed $a_1$ in the case Im$a_1=0$. 
 The left figure shows the positions of the singular points 
 for $a_1=0$, the upper figures shows the evolutions 
 of the singular points  for $a_1>0$, 
 and the lower figures show them as $a_1<0$ is decreasing. }
\label{evo1}
\end{figure}
\end{center}

Here we analyze the vacuum structure of our theory in $N_f=1$ case. 
Let us first discuss the flow of the singular points. 
In the following analysis, the dynamical scale 
 is fixed as $\Lambda_1=(256/27)^{1/6}$. 
The singular points on the $u$-plane are given by the solutions 
 of the cubic polynomial \cite{s-w2},
\begin{eqnarray}
 \Delta=\frac{\Lambda_1^6}{16}
       \left[
        -u^3+m^2u^2+\frac{9}{8}\Lambda_1^3mu
        -\Lambda_1^3m^3-\frac{27}{256}\Lambda_1^6
       \right]\Bigg{|}_{a_1=\frac{m}{\sqrt{2}}} =0 .
\end{eqnarray}
For simplicity, we consider the case $\mbox{Im} a_1=0$ in the following. 
The flow of the singular points is sketched in Fig.~\ref{evo1}. 
For $a_1=0$, there are three singular points 
 at $u_1=-1$, $u_2=\exp(i\pi/3)$ and $u_3=\exp(-i\pi/3)$, respectively. 
These singular points correspond to the appearance of the BPS states 
 with quantum numbers 
 $(n_e,n_m)=(2,1)$, $(1,1)$ and $(0,1)$, respectively. 
In this case, there is non-anomalous ${\bf Z}_3$ symmetry 
 on the moduli space. 
The $(2,1)_1$ dyon point is moving to the left on real $u$-axis, 
 as $a_1$ is increasing. 
The $(1,1)_1$ dyon point and the $(0,1)_0$ monopole point 
 are moving to the right and approaching real $u$-axis, 
 and eventually collide on real $u$-axis 
 for $a_1=\frac{3}{4\sqrt{2}}\Lambda_1$. 
This collision point is called Argyres-Douglas (AD) point \cite{douglas}, 
 at which the two collapsing states are simultaneously massless, 
 and the theory is believed to transform 
 into the superconformal theory. 
After the collision and for $a_1 > \frac{3}{4\sqrt{2}}\Lambda_1$, 
 quantum numbers of two BPS states, 
 $(n_m,n_e)_n=(1,1)_1$ and $(0,1)_0$, change into 
 $(1,0)_1$ and $(0,1)_0$, respectively, 
 due to the conjugation of monodromy \cite{ferrari2}.
As $a_1$ is increasing further, 
 both of the singular points, $u_2$ and $u_3$, 
 are moving to the right on real $u$-axis 
 ($u_2$ approaches the infinity faster than $u_3$). 
On the other hand, as $a_1$ is decreasing from $a_1=0$, 
 the dyon point $u_1$ is moving to the right on real $u$ axis. 
The dyon point $u_2$ and the monopole point $u_3$ 
 are approaching imaginary $u$-axis 
 to the infinities, $u_2 \rightarrow + i \infty$ 
 and  $u_3 \rightarrow - i \infty$, respectively. 

\begin{center}
\begin{figure}[p]
\centerline{\psfig{file=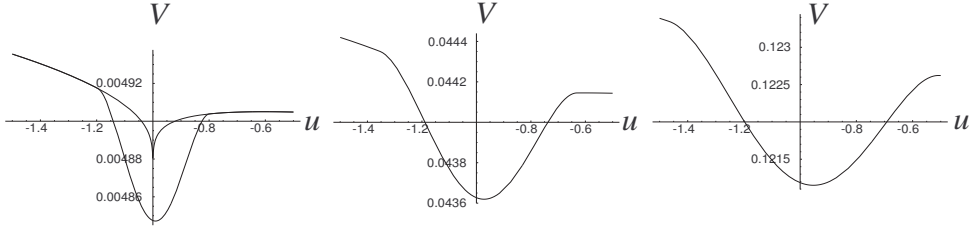,width=17cm}}
\caption{ The effective potential around the $(2,1)_1$ dyon point 
 for $a_1=0$ and $\xi=0.1$, $0.3$, $0.5$ (from left to right) 
 along real $u$-axis.}
\label{dyon-pote}
\end{figure}
\end{center}

Before analyzing the vacuum structure, let us see the dependence 
 of the effective potential on the SUSY breaking parameter $\xi$. 
For $a_1=0$, the effective potential 
 around the dyon singular point $u_1$ is depicted in Fig.~\ref{dyon-pote}. 
 with various values of $\xi$. 
In the left figure, 
 the top figure with the cusp and the bottom figure show 
 the effective potential without and with the dyon condensation, 
 respectively. 
Note that the cusp is smoothed out in the effective potential 
 including the dyon condensation. 
This fact means that the dyon really enjoys the correct degrees 
 of freedom in the effective theory around the singular point. 
We can see that values of the potential minimum 
 and the width of the dyon condensation 
 are controlled by the scale of $\xi$, as expected. 

\begin{center}
\begin{figure}[p]
\centerline{\psfig{file=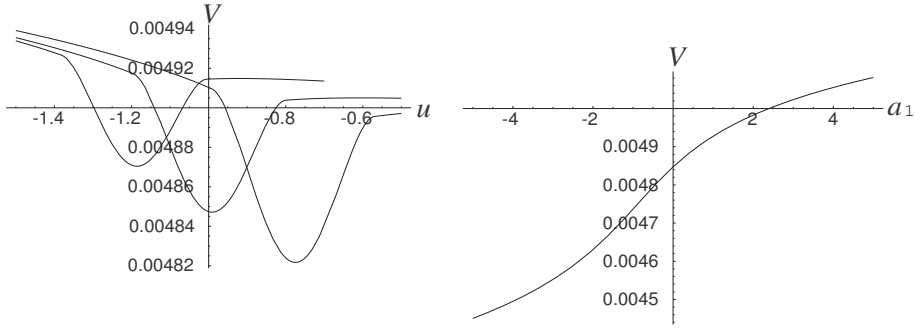,width=16cm}}
\caption{ The left figure shows the effective potential 
  for $a_1=1/8$ (left), $0$ (middle), $-1/8$ (right). 
  The right figure shows the evolution of the potential energy 
  along the flow of the $(2,1)_1$ dyon point. }
\label{dyon-evo}
\end{figure}
\end{center}

Now we investigate the vacuum structure by varying the 
 values of $a_1$. 
In the following analysis, the SUSY breaking parameter is 
 fixed as $\xi=0.1$. 
First, let us see the evolution of the potential energy 
 along the flow of the $(2,1)_1$ dyon point ($u_1$). 
The effective potentials for $a_1=1/8$, $0$ and $-1/8$ from left to right 
 are depicted in Fig.~\ref{dyon-evo}. 
We can check that the potential minimum appears 
 on the singular point for fixed $a_1$. 
The right figure shows the evolution of the potential energy 
 along the flow of the singular point. 
We can find that, as $a_1$ is decreasing, 
 the potential energy is monotonically decreasing. 
Therefore, the effective potential is not bounded from below 
 along the flow of the $(2,1)_1$ dyon point. 

\begin{center}
\begin{figure}[p]
\centerline{\psfig{file=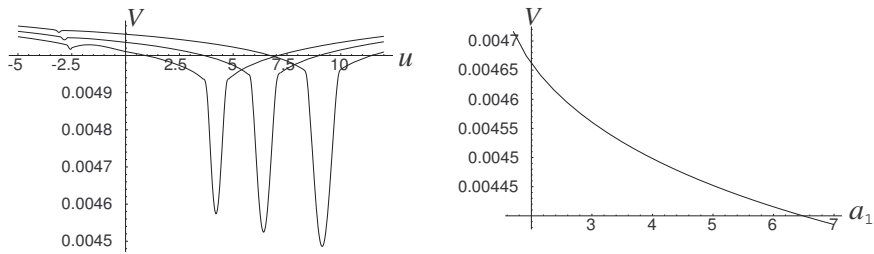,width=15cm}}
\caption{ The left figure shows the effective potential 
  along real $u$-axis for $a_1=\frac{2}{\sqrt{2}}$ (left), 
  $\frac{2.5}{\sqrt{2}}$ (middle) and $\frac{3}{\sqrt{2}}$ (right).
  The right figure shows the evolution of the potential energy 
  along the flow of the quark singular point.}
\label{quark-evo1}
\end{figure}
\end{center}

Next we investigate the effective potential 
 around the other two singular points. 
For $a_1< \frac{3}{4\sqrt{2}}\Lambda_1$, 
 the effective potential has the CP symmetry, and is invariant 
 under the transformation $u \leftrightarrow u^\dagger$.
Hence, the potential energies on these points are degenerate. 
As $a_1$ is increasing, the potential energy is monotonically decreasing 
 toward the AD point ($a_1=\frac{3}{4\sqrt{2}}\Lambda_1$), 
 at which two singular points collide. 
For $a_1 > \frac{3}{4\sqrt{2}}\Lambda_1$,  
 two singular points appear again, 
 the $(0,1)_0$ monopole point and the $(1,0)_1$ quark singular point.  
The effective potential for various values of 
 $a_1 > \frac{3}{4\sqrt{2}}\Lambda_1$ is depicted in Fig.~\ref{quark-evo1}. 
For $a_1 > \frac{3}{4\sqrt{2}}\Lambda_1$, all the singular points 
 are on real $u$-axis. 
From the left figure, we see that there appears the potential minimum 
 only on the quark singular point 
 (the left minimum corresponds to the minimum around $(2,1)_1$ dyon point 
  already depicted in Fig.~\ref{dyon-evo}). 
The $(0,1)_0$ monopole condensation is too small 
 for the effective potential to have its minimum on the monopole point. 
Since, as $a_1$ is increasing further, the potential energy 
 on the quark singular point is monotonically decreasing 
 as depicted in the right figure, 
 we find that the effective potential is not bounded from below 
 along the flow of this singular point. 
\footnote{
 To be correct, there is a possibility 
 that the local minimum may exist near the AD point. 
 However, our description of the effective theory is not applicable 
 around the point, since the condensations of two BPS states 
 are well overlapped. 
 For detailed discussions in this situation, see the next subsection B.}

\begin{center}
\begin{figure}[p]
\centerline{\psfig{file=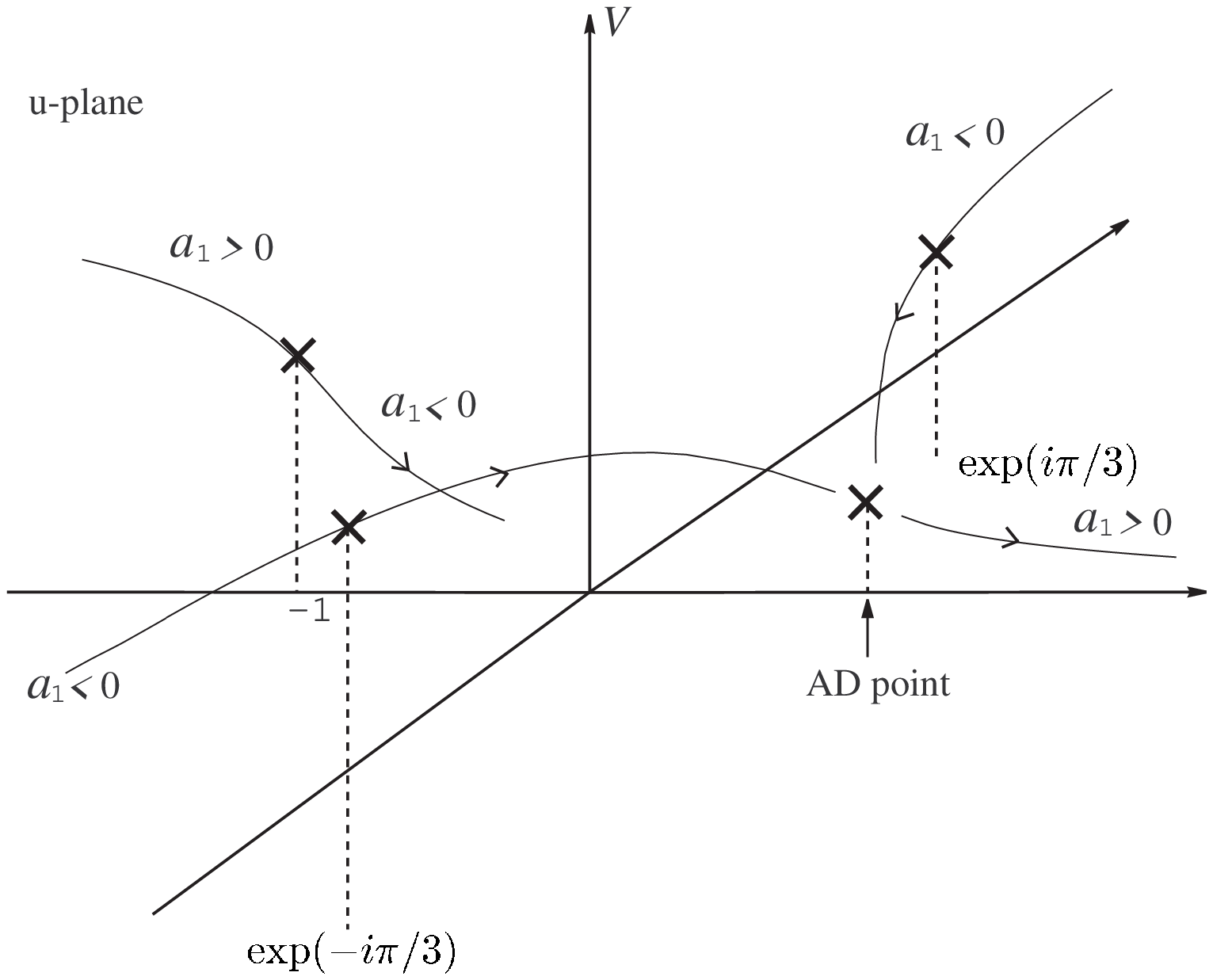,width=14cm}}
\caption{ The evolutions of the potential energies  
  along the flows of the singular points.}
\label{sum1}
\end{figure}
\end{center}

In the case $\mbox{Im} a_1=0$, 
 the evolutions of the potential energies 
 along the flows of the singular points 
 are simultaneously sketched in Fig.~\ref{sum1}. 
The effective potential is found to be the runaway type, 
 and is monotonically decreasing 
 toward the boundary of the defined region of moduli space. 
We can analyze the effective potential for general complex 
 values of $a_1$, and find that the same results come out. 
In conclusion, there is neither well-defined vacuum 
 nor local minimum in the effective theory. 

\subsection{Vacuum Structure in $N_f=2$ case}

\begin{center}
\begin{figure}[p]
\centerline{\psfig{file=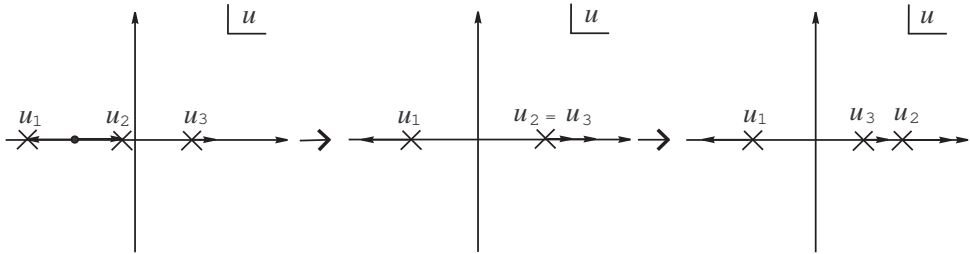,width=16cm}}
\caption{ The flow of the singular points in the case Im$a_1=0$. 
  The figures show the position of the singular points 
  for $a_1=0$, $a_1=\frac{\Lambda_{2}}{2\sqrt{2}}$ 
  and $a_1>\frac{\Lambda_{2}}{2\sqrt{2}}$ 
  from left to right. }
\label{evo2}
\end{figure}
\end{center}

Next, we analyze the vacuum structure for $N_f=2$ case. 
Again, let us first consider the flow of the singular points.  
In the following analysis, the dynamical scale is fixed as 
 $\Lambda_{2}=2\sqrt{2}$. 
In $N_f=2$ case, the discriminant of the algebraic curve 
 can be easily solved such that 
\begin{eqnarray}
u_1=-m\Lambda_2-\frac{\Lambda_2^2}{8}
      \Bigg{|}_{a_1 =\frac{m}{\sqrt{2}}}, \; 
u_2=m\Lambda_2-\frac{\Lambda_2^2}{8}
      \Bigg{|}_{a_1 =\frac{m}{\sqrt{2}}}, \;
u_3=m^2+\frac{\Lambda_2^2}{8}
      \Bigg{|}_{a_1 =\frac{m}{\sqrt{2}}}. \label{descriminat}  
\end{eqnarray}
We investigate the case $\mbox{Im} a_1 =0$, for simplicity.  
The flow of the singular points is sketched in Fig.~\ref{evo2}. 
For $a_1=0$, the singular points appear 
 at $u_1=u_2=-1$ and $u_3=1$. 
Here, at $u=-1$, two singular points degenerate. 
For non-zero $a_1>0 $, 
\footnote{ 
 We consider only the case $a_1 > 0$, 
 since the result for $a_1 < 0$ can be obtained 
 by exchanging $u_1 \leftrightarrow u_2$,
 as be seen from the first two equations in Eq.~(\ref{descriminat}).} 
 this singular point splits into two singular points $u_1$ and $u_2$,
 which correspond to the BPS states with quantum numbers 
 $(1,1)_{-1}$ and $(1,1)_1$, respectively.
As $a_1$ is increasing, these singular points, $u_1$ and $u_2$, 
 are moving to the left and the right on real $u$-axis, respectively.  
Two singular points, $u_2$ and $u_3$, collide and degenerate 
 at the AD point ($u=\frac{3\Lambda_2^2}{8}$) for 
 $a_1= \frac{\Lambda_2}{2 \sqrt{2}}$. 
As $a_1$ is increasing further, there appear two singular points 
 $u_2$ and $u_3$ again, 
 and quantum numbers of the corresponding BPS states, 
 $(1,1)_1$ and $(0,1)_0$, change into $(1,0)_1$ and $(1,-1)_1$, 
 respectively. 
The singular point $u_2$ is moving to the right faster than $u_3$. 

\begin{center}
\begin{figure}[p]
\centerline{\psfig{file=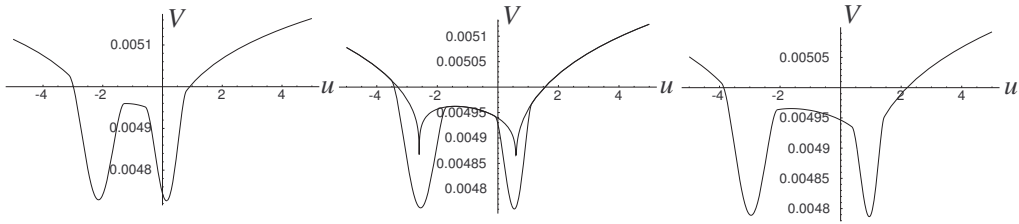,width=17cm}}
\caption{ The effective potential for $a_1=0.3$ (left), 
 $a_1=0.4$ (middle) and $a_1= 0.5$ (right).} 
\label{fix}
\end{figure}
\end{center}

\begin{center}
\begin{figure}[p]
\centerline{\psfig{file=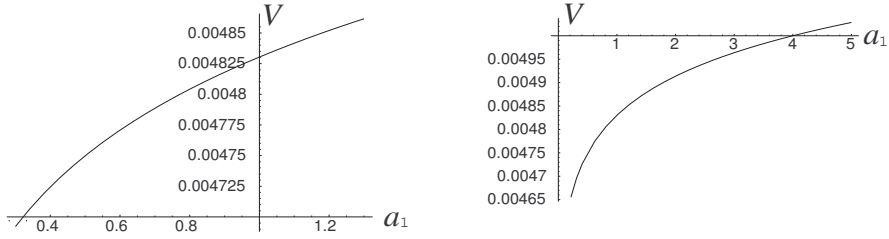,width=16cm}}
\caption{ The evolutions of the potential energies 
 along the flows of the $(1,1)_{-1}$ dyon point (left) 
 and the $(1,1)_{1}$ dyon point (right). }
\label{flavor2-2}
\end{figure}
\end{center}

We investigate the vacuum structure by varying the values of $a_1$. 
For $ 0 < a_1 < \frac{\Lambda_2}{2 \sqrt{2}}$, 
 the effective potential is plotted in Fig.~\ref{fix}. 
While there appear the potential minima 
 at two singular points $u_1$ and $u_2$, 
 the monopole condensation is too small for the potential 
 to have a minimum at the singular point $u_3$. 
The top and bottom figures in the middle show the effective potential 
 without and with the dyon condensations, respectively. 
The cusps are smoothed out in the bottom figure, as in the case $N_f=1$. 
The evolutions of the potential energies on the singular points 
 $u_2$ and $u_3$ are depicted in Fig.~\ref{flavor2-2}.  
We find that both of them are decreasing toward the point $a_1=0$, 
 and thus the effective potential is bounded from below,  
 at least, along real $u$-axis. 

\begin{center}
\begin{figure}[p]
\centerline{\psfig{file=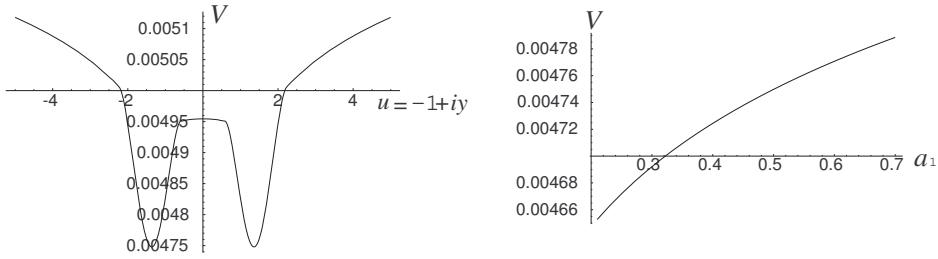,width=16cm}}
\caption{ The effective potential (left)
 for $a_1=i\frac{\sqrt{2}}{4}$ 
 along imaginary $u$-axis with Re$u=-1$, 
 and the evolution of the potential energy (right) 
 along the flow of the singular point $u_1$ 
 in the case Re$a_1=0$.} 
\label{flavor2-4}
\end{figure}
\end{center}

Next, we examine whether the effective potential is bounded 
 in all the directions for general complex $a_1$ values. 
As an example, let us consider the case $\mbox{Re}a_1=0$. 
For $a_1 \neq 0$, the two singular points $u_1$ and $u_2$  
 appear on the imaginary $u$-axis with $\mbox{Re} u =-1$. 
The effective potential is depicted in Fig.~\ref{flavor2-4} 
 along this axis for $a_1= i \frac{\sqrt{2}}{4}$. 
There appear two potential minima at the singular points. 
The right figure shows the evolution of the potential energy 
 along the flow of the singular point $u_1$, 
\footnote{ 
 Two potential minima for fixed $a_1$ are degenerate, 
 since the effective potential has the CP symmetry 
 under the exchange $u \leftrightarrow u^\dagger$ 
 in the case $\mbox{Re}a_1=0$.} 
and we find that the effective potential is also bounded in this direction. 
We can check that the effective potential is bounded from below 
 for all the values of small $|a_1|$. 
Therefore, the effective potential seems 
 to have the local minimum at the points $u=-1$ and $a_1=0$.   

\begin{center}
\begin{figure}[p]
\centerline{\psfig{file=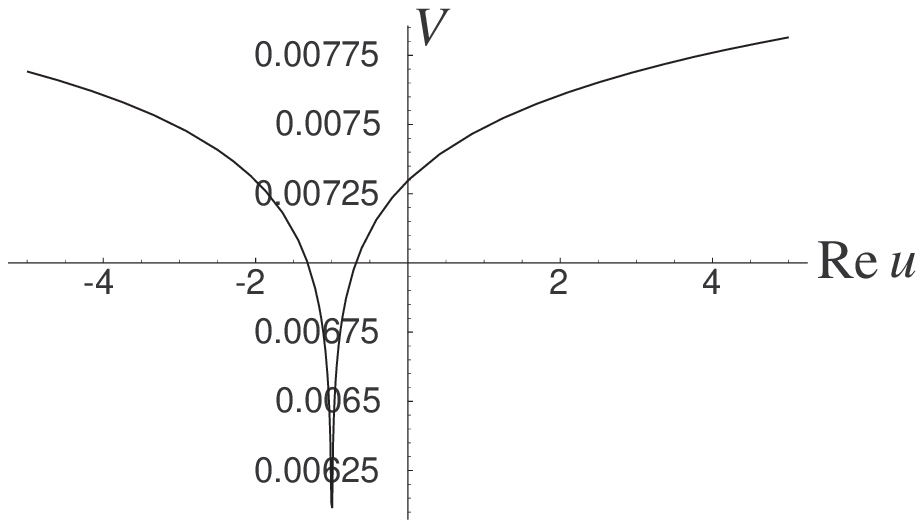,width=16cm}}
\caption{The effective potential without the 
 contribution of the dyon condensations 
 in the limit $a_1\rightarrow 0$.}
\label{flavor2-5}
\end{figure}
\end{center}

However, note that our description is not applicable  for small $|a_1|$, 
 since the condensations of two dyon states are going to 
 overlap with each other (see Fig.~\ref{fix}). 
Unfortunately, we have no knowledge about the correct description 
 of the effective theory in this situation. 
Nevertheless, we conclude that there must appear the local minimum 
 with broken SUSY in the limit $a_1 \rightarrow 0$ 
 from the result in the following. 
In this limit, the effective potential without the dyon condensations 
 is depicted in Fig.~\ref{flavor2-5}.
We can find that there appears the potential minimum at $u=-1$, 
 and the value of the effective potential on the cusp 
 is non-zero, $V \sim 0.0061 >0 $. 
If we had the correct description of the effective theory for $a_1$=0, 
 this cusp might be smoothed out.  
However, there is no reason that SUSY is restored at $u=-1$, 
 because the correct effective theory must have no singularity 
 for the Kahler metric.  
Therefore, there is the promising possibility 
 of the appearance of the local minimum with broken SUSY 
 at $u=-1$ and $a_1=0$.  

Finally, let us get back to the case $\mbox{Im} a_1=0$. 
For $a_1 >  \frac{\Lambda_2}{2 \sqrt{2}}$,  
 the effective potential has two minima 
 only at two singular points $u_1$ and $u_2$. 
The monopole condensation is too small for the effective potential 
 to have a minimum at $u_3$. 
The plot of the effective potential is similar to Fig.~\ref{quark-evo1}. 
While the evolution of the potential energy along the singular point $u_1$ 
 is the same as for $0 < a_1 < \frac{\Lambda_2}{2 \sqrt{2}}$, 
 the potential energy on the quark singular point $u_2$ 
 is monotonically decreasing, as $a_1$ is increasing. 
Thus, there is a runaway direction 
 along the flow of the quark singular point.  
We can find the same global structure 
 along the flow of the quark singular point  
 for general complex $a_1$ values. 
   
\begin{center}
\begin{figure}[p]
\centerline{\psfig{file=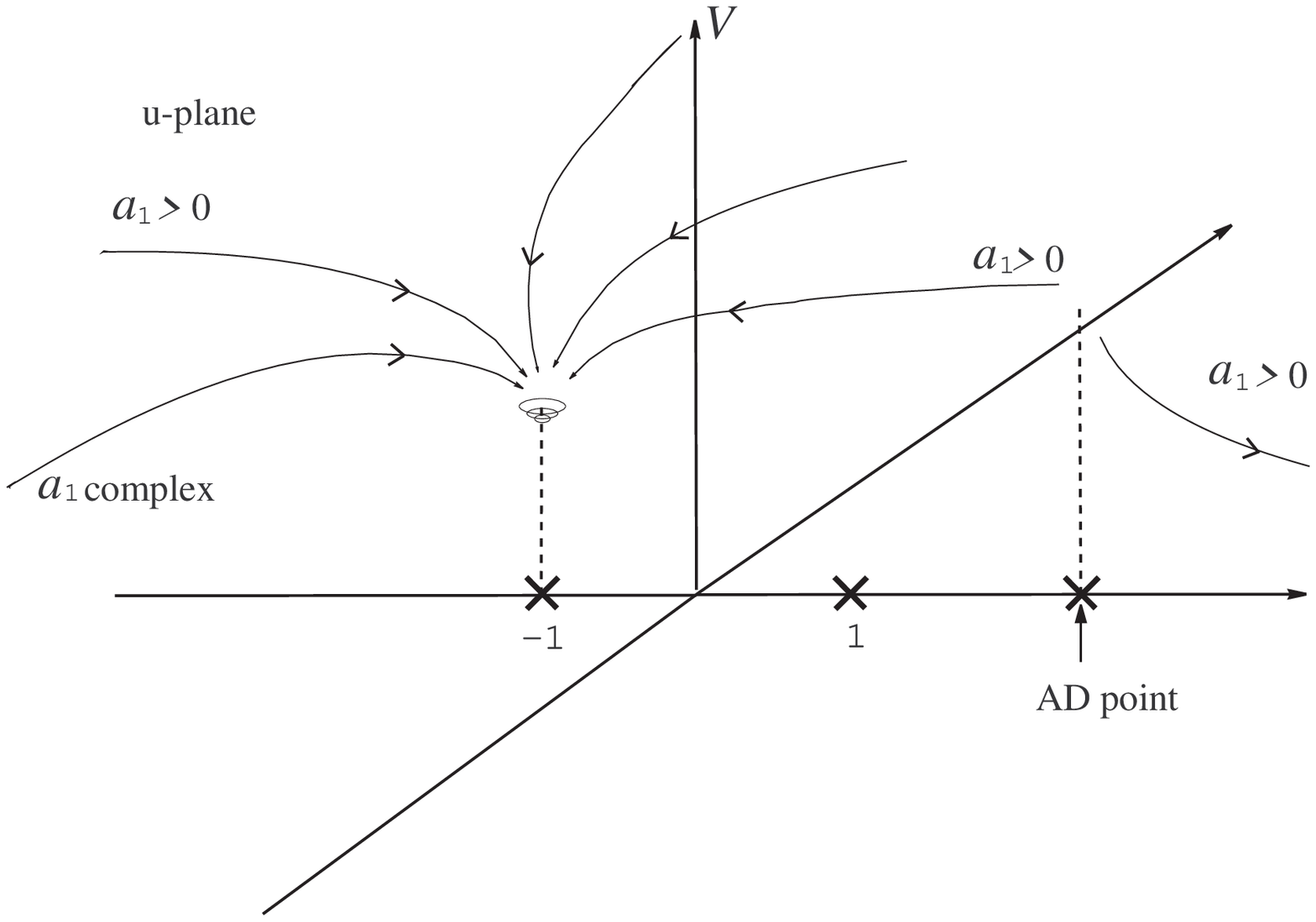,width=16cm}}
\caption{The evolutions of the potential energies 
 along the flows of the singular points.} 
\label{sum2}
\end{figure}
\end{center}

The evolutions of the potential energies 
 along the flows of the singular points 
 are simultaneously sketched in Fig.~\ref{sum2}. 
The global structure of the effective potential 
 is the same as in the case $N_f=1$, namely, the runaway type. 
However, we find the promising possibility that 
 there exists the local minimum with broken SUSY in the theory. 
Since there is no well-defined vacuum on the runaway direction, 
 this minimum with broken SUSY is  
 the unique and promising candidate for the vacuum in the theory. 
Unfortunately, we have no knowledge of the correct description 
 about the effective theory around the degenerate dyon point.

\section{Conclusion} 

We analyzed the vacuum structure of spontaneously broken 
 ${\cal N}=2$ SUSY gauge theory with the Fayet-Iliopoulos term. 
Our theory is based on the gauge group $SU(2) \times U(1)$ 
 with $N_f=1,2$ massless quark hypermultiplets 
 having the same $U(1)$ charges. 
The $U(1)$ gauge interaction was necessary introduced 
 to include the Fayet-Iliopoulos term in the theory. 
It was shown that there are degenerate vacua 
 in the classical potential even in the absence of SUSY. 
This degeneracy is expected to be smoothed out, 
 once quantum corrections are taken into account. 
Then, we investigated the effective potential, 
 and analyzed the the vacuum structure of the theory. 

The effective action was formulated up to the leading order 
 for the SUSY breaking parameter $\xi$. 
In our framework, the $U(1)$ gauge interaction part 
 was treated as the cut-off theory 
 under the assumption that the $U(1)$ gauge interaction 
 has no effect on the $SU(2)$ gauge dynamics. 
Thus, the moduli space in the theory was restricted 
 within the region smaller than the cut-off scale, the Landau pole. 
Considering the monodromy transformation, we found that 
 the prepotential consistent with the assumption 
 was the same as the one in SQCD with massive quark hypermultiplets. 

The effective potential was the function of the moduli parameters. 
Examining the minimum of the effective potential, 
 we found that the singular points on the moduli space 
 were energetically favored, 
 because of the condensations of the light BPS states. 
The singular points on the $u$-plane flowed 
 according to the values of the moduli parameter $a_1$. 
Thus, we analyzed the effective potential 
 along the flows of the singular points, 
 and examined which point was energetically favored 
 on the line of the flow. 
 
In $N_f=1$ case, the effective potential was found 
 to be the runaway type and monotonically decreasing 
 toward the ultraviolet region in the moduli space. 
We observed that there was neither well-defined vacuum 
 nor the local minimum in this case. 
In $N_f=2$ case, there was also the runaway direction 
 along the flow of the quark singular point, 
 and the global structure was the same as in $N_f=1$ case. 
However, we found the promising possibility that 
 the local minimum with broken SUSY appears at the degenerate dyon point. 
Therefore, this point is the promising candidate 
 for the well-defined vacuum. 
Unfortunately, we have no knowledge about the correct description 
 of the effective theory around the degenerate singular point, 
 since the condensations of two BPS states well overlap there. 

The difference of our result from that in SQED is worth noticing. 
In SQED, the potential minimum appears at the massless singular point, 
 and SUSY is formally restored there, 
 because of the singularity of the Kahler metric.  
On the other hand, in our case with $N_f=2$ hypermultiplets, 
 we found that the value of the potential minimum 
 was non-zero, and thus SUSY was really broken there 
 (at least in analysis of the effective potential 
  without the dyon condensations). 
This result means that the singularity of the Kahler metric 
 is removed by the effect of the $SU(2)$ gauge dynamics 
 in the infrared region on the moduli space. 

Finally we comment on the possibility 
 that the theory has the global minimum. 
Note that the behavior of the effective potential changes  
 according to the number of flavors $N_f$, 
 since both of the periods and the effective couplings depend on $N_f$. 
Indeed, we observed that the structures of the effective potentials 
 were different in two cases $N_f=1,2$. 
If we consider the extended version of our theory 
 with $N_f=3,4$ quark hypermultiplets, 
 we may find the global minimum in the theory. 
Although this extension makes our analysis 
 much more complicated and difficult, it will be interesting.

\acknowledgments
One of the authors (M. A.) is grateful to Marcos Marino 
 for his useful advice for technical details in our analysis. 
We would like to thank Noriaki Kitazawa and Satoru Saito
 for useful discussions and comments. 

\appendix
\section{}

In this appendix, we demonstrate the derivations 
 of the integrals $I_i^{(j)}$ 
 in Eqs.~(\ref{eq:formula1})-(\ref{eq:formula3}). 
The complete elliptic integrals are given as follows.
\begin{eqnarray}
K(k)&=&\int_0^1 \frac{dx}
     {\left[(1-x^2)(1-k^2x^2)\right]^{1/2}}, 
       \\ \nonumber
E(k)&=&\int_0^1 dx\left(\frac{1-k^2x^2}{1-x^2}
     \right)^{1/2}, 
       \\ \nonumber
\Pi_1(\nu,k)&=&\int_0^1\frac{dx}
     {[(1-x^2)(1-k^2x^2)]^{1/2}(1+\nu x^2)}.
\end{eqnarray}
The integrals $I_i^{(j)}$ are described in terms of 
 the above complete elliptic integrals 
 through some steps of changing variables in the integrations. 

First, demonstrate the derivation of Eq.~(\ref{eq:formula1}). 
\begin{eqnarray}
I_1^{(1)}&=&\int_{e_2}^{e_3}\frac{dX}{Y} 
            \nonumber \\
         &=&\int_{e_2}^{e_3}\frac{dX}
            {\sqrt{4(X-e_1)(X-e_2)(X-e_3)}}
            \nonumber \\
         &=&\frac{i}{2\sqrt{e_2-e_1}} \int_0^{k^{\prime 2}}
            \frac{dt}{[t (t-1)(t -k^{\prime 2})]^{1/2}},
\end{eqnarray}
where we changed the variable $X$ by $t =-\frac{X-e_2}{e_2-e_1}$.
Further, changing the variable $t$ by 
 $t=1+k+\frac{1}{\zeta-\frac{1}{2k}}$ 
 and rescaling $\zeta$ as $x=2k\frac{1+k}{1-k}\zeta$, we obtain 
\begin{eqnarray}
I_1^{(1)}=\frac{i}{\sqrt{e_2-e_1}}\frac{2}{1+k}
          K\left(\frac{1-k}{1+k}\right).
\end{eqnarray}
Using the relation 
\begin{eqnarray}
\frac{2}{1+k}K\left(\frac{1-k}{1+k}\right)
        =K(k^\prime), \label{eq:ellip1}
\end{eqnarray}
 we obtain Eq.~(\ref{eq:formula1}). 
Repeating the same steps for the integral $I_2^{(1)}$, we obtain 
\begin{eqnarray}
I_2^{(1)}&=&\int_{e_2}^{e_3}\frac{XdX}{Y} 
            \nonumber \\ 
         &=&\frac{i}{\sqrt{e_2-e_1}} \frac{2}{1+k} 
            \Bigg{[}
            \{e_2-(e_2-e_1)(1+k)\}
            K\left(\frac{1-k}{1+k}\right)
            \nonumber \\ 
         & &\left.
            +2k(e_2-e_1)
            \Pi_1
            \left(-\left(\frac{1-k}{1+k}\right)^2, 
            \frac{1-k}{1+k}\right)
            \right].
\end{eqnarray}
Using Eq.~(\ref{eq:ellip1}) and the following relations,  
\begin{eqnarray}
(1-\bar{k}^2)\Pi(-\bar{k}^2,\bar{k})=E(\bar{k}), \; 
(1+k)E\left(\frac{1-k}{1+k}\right) = E(k^\prime)+k K(k^\prime), 
 \label{eq:relations} 
\end{eqnarray}
 where $\bar{k}=\frac{1-k}{1+k}$, we obtain Eq.~(\ref{eq:formula2}). 
Finally, for $I_3^{(1)}$, the same steps lead to 
\begin{eqnarray}
I_3^{(1)}(c)&=&\int_{e_2}^{e_3}\frac{dX}{Y(X-c)} \nonumber \\ 
&=&\frac{-i}{(e_2-e_1)^{\frac{3}{2}}}
   \frac{1}{\tilde{c}+k}\frac{1}{1+k}  \nonumber \\ 
& \times &  \int_{-1}^1\frac{dx}
    {\left[\left(1-\left(
             \frac{1-k}{1+k}\right)^2
             x^2\right)(1-x^2)
    \right]^{\frac{1}{2}}}
\left[1-\frac{2k}{\tilde{c}+k}
          \frac{1}{\frac{1-k}{1+k}x- \frac{\tilde{c}-k}
                     {\tilde{c}+k}  } \right], 
\end{eqnarray}
where $\tilde{c}=\frac{c-e_1}{e_2-e_1}$. 
 Using the relations Eq.~(\ref{eq:ellip1}), 
 we obtain Eq.~(\ref{eq:formula3}).

\section{} 

In this subsection, we show the derivations 
 of the effective couplings in term of the Weierstrass functions. 
It is convenient to introduce the uniformization variable $z$ 
 through the map with the Weierstrass $\wp$ function, 
\begin{eqnarray}
 (\wp(z),\wp^\prime(z))=(X,Y).\label{eq:map}
\end{eqnarray}
Using this map, the half period $\omega_i/2$ 
 is mapped into the root $e_i=\wp(\omega_i/2)$ 
 ($\omega_3=\omega_1+\omega_2$). 
The inverse map is defined as 
\begin{eqnarray}
 z=\Psi^{-1}(x_0)=\int_{x_0}^\infty\frac{dX}{Y}
 =-\frac{1}{\sqrt{e_2-e_1}}F(\phi,k),
\end{eqnarray}
where we changed the integration variable $X$ 
 by $t^2=(e_2-e_1)/(X-e_1)$, 
 and $F(\phi,k)$ is the incomplete elliptic integral given by
\begin{eqnarray}
F(\phi,k)=\int_0^{\sin\phi}\frac{dt}
          {[(1-t^2)(1-k^2t^2)]^{1/2}} ; \; \; 
       \sin^2\phi=\frac{e_2-e_1}{x_0-e_1}.
\end{eqnarray}

We derive the effective couplings, $\tau_{12}$ and $\tau_{11}$, 
 by using the map of Eq.~(\ref{eq:map}). 
The effective coupling $\tau_{12}$ is described by 
\begin{eqnarray}
\tau_{12}=\frac{\partial a_{2D}}{\partial a_1}
          \Bigg{|}_{a_2}
         =\frac{\partial a_{2D}}{\partial a_1}
           \Bigg{|}_u
          -\tau_{22}
          \frac{\partial a_{2}}{\partial a_1}
           \Bigg{|}_u. \label{eq:tau12a}
\end{eqnarray}
The partial derivative of the periods $a_{2D}$ and $a_2$ 
 with respect to $a_1$ can be calculated by using 
 Eqs.~(\ref{eq:period})-(\ref{eq:f2}) as 
\begin{eqnarray}
 \frac{\partial a_{2i}}{\partial a_1}\Bigg{|}_u
  = \oint_{\alpha_i}
       \frac{\partial \lambda_{SW}}{\partial a_1}\Bigg{|}_u
  = Q^{(N_f)}(a_1,~\Lambda_{N_f})
       \int_{e_j}^{e_3}\frac{dX}{2Y(X-c)} \;~~(i\neq j), 
\end{eqnarray}
where the coefficient $Q^{(N_f)}$ is given by
\begin{eqnarray}
Q^{(N_f)}(a_1,\Lambda_{N_f}) 
 =-\frac{N_f(\sqrt{2}a_1)^{N_f-1}\Lambda_{N_f}^{4-N_f}}{16\pi}.
\end{eqnarray}
Using the map of Eq.~(\ref{eq:map}), the integral can be described as 
\begin{eqnarray}
 \frac{\partial a_{2i}}{\partial a_1}\Bigg{|}_u 
  & = &Q^{(N_f)}
       \int_{\omega_j}^{\omega_3}\frac{dz}{2(\wp(z)-\wp(z_0))}
       \nonumber \\
  & = &\frac{Q^{(N_f)}}{2}
       \frac{1}{\wp^\prime(z_0)}
       \left(\log\frac{\sigma(z-z_0)}{\sigma(z+z_0)}
             +2z\zeta(z_0)
       \right),
\end{eqnarray}
where $\wp(z_0)=c$, $\zeta(z)$ is the Weierstrass zeta function, 
 and we used the definition of the Weierstrass sigma function, 
 $\zeta(z)=\frac{d}{dz}\log\sigma(z)$, and the relation
\begin{eqnarray}
\frac{\wp^\prime(z_0)}{\wp(z)-\wp(z_0)}
        =\zeta(z-z_0)-\zeta(z+z_0)+2\zeta(z_0). 
\end{eqnarray}
Taking into account that $Y$ corresponds to $\wp^\prime(z)$ 
 under the map of Eq.~(\ref{eq:map}),
 the pole $\wp^\prime(z_0)$ can be easily obtained as  
\begin{eqnarray}
\wp^\prime(z_0)^2=-\left(\frac{N_f2^{(N_f-1)/2}
                          \Lambda_{N_f}^{4-N_f}}{32}
                   \right)^2.
\end{eqnarray}
Using the pseudo periodicity of the Weierstrass sigma function, 
\begin{eqnarray}
\sigma(z_0+\omega_i)=
 -\sigma(z_0)\exp\left(2\zeta
  \left(\frac{\omega_i}{2}\right)
       \left(z_0+\frac{1}{2}\omega_i\right)
  \right),
\end{eqnarray}
we obtain
\begin{eqnarray}
\frac{\partial a_{2i}}{\partial a_1}\Bigg{|}_u
 =-\frac{1}{\pi i}
      \left[\omega_i\zeta(z_0)
            -2z_0\zeta\left(\frac{\omega_i}{2}\right)
      \right].\label{eq:anobibun}
\end{eqnarray}
The zeta function at half period can be described by the integral 
 representations such as
\begin{eqnarray}
 \zeta\left(\frac{\omega_i}{2}\right)=-I_2^{(i)}.\label{eq:int}
\end{eqnarray}
Substituting Eq.~(\ref{eq:anobibun}) into Eq.~(\ref{eq:tau12a}) 
 and using the Legendre relation 
\begin{eqnarray}
\omega_1\zeta\left(\frac{\omega_2}{2}\right)
 -\omega_2\zeta\left(\frac{\omega_1}{2}\right)
 =i\pi,
\end{eqnarray}
we finally obtain 
\begin{eqnarray}
\tau_{12}=-\frac{2z_0}{\omega_2}.
\end{eqnarray}

Next we derive the effective coupling $\tau_{11}$, 
 which is given by differentiating $a_{1D}$ of Eq.~(\ref{eq:mass}) 
 with respect to $a_1$ with $a_2$ fixed such as 
\begin{eqnarray}
\tau_{11} =-
           \int_{x_n^-}^{x_n^+}
           \left[\frac{\partial \lambda_{ SW}} {\partial u}
                 \Bigg{|}_{a_1}            
                  \frac{\partial u}
                  {\partial a_1}
                  \Bigg{|}_{a_2}
                +\frac{\partial \lambda_{SW}}
                      {\partial a_1}
                 \Bigg{|}_u
           \right]
           +\frac{\partial \tilde{C}}{\partial a_1}. \label{eq:pretau11}
\end{eqnarray}
The integral can be evaluated by using the map (\ref{eq:map}).
Although the integral contains divergence, 
 it can be regularized by using the freedom 
 of the integration constant $\tilde{C}$. 
Let us demonstrate this regularization 
 by introducing the regularization parameter $\epsilon$ as follows. 
\begin{eqnarray}
\tau_{11}&=&-\int_{x_0^-+\epsilon}^{x_0^++\epsilon}
             \left[\frac{\partial \lambda_{ SW}} {\partial u}
                 \Bigg{|}_{a_1}
                 \left(-\frac{\partial u}{\partial a_2}\Bigg{|}_{a_1}
                        \frac{\partial a_2}{\partial a_1}\Bigg{|}_{a_2} 
                 \right)
                 +\frac{\partial \lambda_{SW}}
                      {\partial a_1}
                 \Bigg{|}_u
             \right]
             +\frac{\partial \tilde{C}}{\partial a_1}
             \nonumber \\
         &=&-\int_{-z_0+\epsilon}^{z_0+\epsilon}dz
             \left[-\frac{1}{\pi i\omega_2}            
                 \left(\omega_2\zeta(z_0)
                  -2z_0\zeta\left(\frac{\omega_2}{2}\right)
                 \right)
                 +\frac{Q^{(N_f)}}{4(\wp(z)-\wp(z_0))}
             \right]
             +\frac{\partial \tilde{C}}{\partial a_1}
             \nonumber \\
         &=&-\frac{1}{\pi i}\left(\log\sigma(2z_0)
              -\frac{4z_0^2}{\omega_2}\zeta
               \left(\frac{\omega_2}{2}\right)\right)
              +\frac{1}{\pi}\log\sigma(\epsilon)+\frac{1}{2}
              +\frac{\partial \tilde{C}}{\partial a_1}.
\end{eqnarray}
The divergence part, $\log\sigma(\epsilon)$, can be subtracted 
 by taking the integration constant such that 
 $\tilde{C}=Ca_1-\frac{a_1}{2}-\frac{a_1}{\pi}\log\sigma(\epsilon)$, 
 and we finally obtain Eq.~(\ref{eq:tau11}) 
 with the relation of Eq.~(\ref{eq:int}).



%
%
%
%
%
%
%
%
%
%
%
%
%
%
\end{document}